\documentclass[useAMS,usenatbib]{mn2e}

\usepackage[dvips]{graphicx}
\usepackage[section]{placeins}
\usepackage{amssymb}

\title[Dynamics of galaxies]
  {Dynamical segregation of galaxies in groups and clusters}

\author[Lares, Lambas \& S\'anchez]
{ M. Lares, D. G. Lambas, A. G. S\'anchez\\
Grupo de Investigaciones en Astronom\'{\i}a Te\'{o}rica 
    y Experimental (IATE), Observatorio Astron\'omico de 
    C\'{o}rdoba, UNC, Argentina.\\
Consejo Nacional de Investigaciones Cient\'{\i}ficas y Tecnol\'ogicas.
    (CONICET), Argentina.\\}

\begin{document}
%\date{Released 2002 Xxxxx XX}

\maketitle

\begin{abstract}

	We have performed a systematic analysis of the dynamics
	of different galaxy populations in galaxy groups from the 2dFGRS.
	For this purpose we have combined all the groups into a single
	system, where velocities $v$ and radius $r$ are expressed
	adimensionally. We have used several methods to compare the
	distributions of relative velocities of galaxies with respect to
	the group centre for samples selected according to their spectral
	type (as defined by Madgwick et al., 2002), $b_j$ band luminosity
	and $B\!-\!R$ colour index.  We have found strong segregation
	effects: spectral type I objects show a statistically narrower
	velocity distribution than that of galaxies with a substantial
	star formation activity (type II-IV). Similarly, the same
	behavior is observed for galaxies with colour index
	$B\!-\!R\!>\!1$ compared to galaxies with $B\!-\!R\!<\!1$.
	Bright ($M_{b}<-19$) and faint ($M_{b}>-19$) galaxies show the
	same segregation. It is not important once the sample is
	restricted to a given spectral type.  These effects are
	particularly important in the central region ($R_p<0.5\;R_{vir}$)
	and do not have a strong dependence on the mass of the parent
	group. These trends show a strong correlation between the
	dynamics of galaxies in groups and star formation rate reflected
	both by spectral type and by colour index.  

\end{abstract}


\section{Introduction}

Galaxy properties can be affected by several mechanisms in groups
or clusters. The fact that different galaxies can be modified to
a different extent, could give rise to observable segregational
effects. By studying these effects, we may obtain valuable
information on the way in which these mechanisms act on galaxies
and drive their evolution.

The morphology density relation \citep{oelmer,dress80,andreon}
is the best known segregational effect. Early type
galaxies are more concentrated in denser regions, and lie
closer to the centres of the clusters than late type galaxies.
More recently, the clustering properties of galaxies have been
found to be dependent on the characteristics of spectral features
\citep{julian,mardom,biviano,madg03c}, and luminosity 
\citep{benoist,norberg01,norberg02,stein,adami,girardi}.

Several mechanisms have been proposed to explain galaxy
transformations. Their relevance are quite different according
to the environmental conditions \citep{balogh}
and so, their importance depends
on the mass of the clusters, and perhaps on the history of galaxy
clustering \citep{gnedin.b}.
Some effects are more effective in dense regions like rich clusters,
whereas in groups of galaxies other mechanisms play the most 
important role. 

Ram pressure \citep{gunn-gott} can inhibit star formation by
exhausting the gas present in galaxies that move fast in the
intergalactic medium of rich clusters.  Similarly, galaxy harassment
\citep{moore} can produce significant changes in the star formation
rate of a galaxy.  These effects are not expected to be important in
poor clusters or groups, where the velocity dispersion is lower,
instead, effects such as mergers or tidal interactions can be dominant
in these environments \citep{gnedin.a}.

Besides affecting galaxy properties, such as star
formation, luminosity and colour index, some of the physical
processes listed above may also produce changes on the dynamics
of the galaxy with respect to the cluster centre.  In turn, the
efficiency of some of these mechanisms to produce significant
changes on a galaxy, depends on its dynamical behavior.  For
example, the effects of galaxy interactions are stronger for
galaxies moving slowly with respect to the cluster centre.  This
suggests that the dynamical properties of galaxies in groups and
clusters may be related with the star formation efficiency,
colours or luminosities of galaxies\citep{menciff,moore2}.
Segregation effects of galaxy velocities in clusters are
predicted theoretically \citep{menciff,gnedin.b}, in
semi--analytical models \citep{menci}, and has been reported in
rich clusters \citep{sodre}.

The relation between the dynamical properties and the luminosity
of a galaxy has been observed in rich clusters, by e.g.
\citet{whitmore,adami} and \citet{stein}, who find evidence
for velocity segregation. 
In agreement with these findings, 
theoretical studies \citep{fyf} and 
numerical simulations \citep{yepes} show similar trends.

However, it is not clear how to interpret these results.
Some authors propose different orbit shapes for galaxies with
different morphologies or luminosities.  In this scheme,
early--type galaxies have quasi--isotropic orbits, while
late--type galaxies move in nearly radial orbits
\citep{biviano77,adami}. However, other models have been proposed
that contradict this statement \citep{amelia}.

Theoretical works predict virialized systems with a Maxwellian
velocity distribution \citep{saslaw,ueda} so it has been proposed
that early and late-type galaxies have Gaussian velocity
distributions.  However, observations in rich clusters do not
support these hypothesis (e.g. Colles \& Dunn, 1996)

The purpose of this paper is to explore for a possible difference
in the dynamical behavior of galaxies with different spectral
types, luminosities and colour indexes.  The outline of this
paper is as follows.  In section 2 we describe the data sample
used in our work, and in section 3, the method used in our
analysis.  Section 4 presents the results of our search for
velocity segregation in spectral type, luminosity and colour.
Finally, in section 5 we present a discussion of our results and
future perspectives.

\section{The data}

\begin{figure}
  \includegraphics[width=\columnwidth]{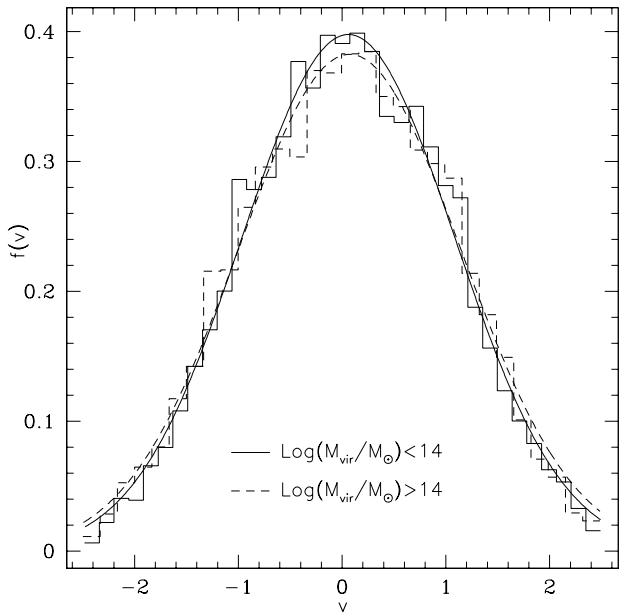}
  \label{gaussmass}
  \caption{Normalized velocity distribution functions of relative
           velocities for two samples of galaxies selected according
           to membership to low ($<10^{14}\;M_{\odot}$) and high
           ($>10^{14}\;M_{\odot}$) parent cluster virial mass.
           Smoothed curves are Gaussian fits centered in $v=0$.}
\end{figure}%

In order to analyse the dynamics of different galaxy populations
in galaxy systems, we have carried out a systematic search for
segregation effects of galaxies in velocity space. We have used a
group catalogue constructed from the final version of the 2dF
Galaxy Redshift Survey \citep{colles}, using the same technique
used by \citet{manuelyz} to construct the group catalogue of the
2dFGRS 100K release \citep{folkes}.  This sample comprises 40978
galaxies in 5568 groups. Virial mass, velocity dispersion and
virial radius have been determined for the groups in this sample.

Principal component analysis technique has been applied to all galaxies
in the 2dFGRS by \citet{madg02}. This technique allows to obtain the
maximum possible spectral information with a minimum set of parameters,
and so it offers a quantitative and efficient way to classify galaxies
with a spectral index $\eta$. This index has a clear physical
interpretation, since 
it is strongly related to the galaxy morphology \citep{madg03a},
and correlates with the equivalent width of the
H$\alpha$ line and with the star birthrate parameter \citep{madg03b}.
These spectral parameters measure the strength of absorption features
by stars and ISM, and the strength of nebular emission features, making
possible to obtain an idea of the relative contributions of different
populations of stars. Negative values of $\eta$ correspond to non
star-forming galaxies, usually early type galaxies, whereas large
values imply star formation features in the integrated spectra,
typical of late type galaxies. \citet{madg02} defined 4 spectral types
based on the shape of the distribution of $\eta$ for galaxies in the
2dFGRS.  Type I comprises all galaxies with $\eta<-1.4$; type II,
galaxies with $-1.4<\eta<1.1$; type III, galaxies with $1.1<\eta<3.1$
and type IV all galaxies with $\eta>3.1$.

The 2dFGRS also contains photometric information in the APM $b_j$ band
and in the super-cosmos $b$ and $r$ bands. 
This is a useful tool to study the dependence of a possible velocity 
segregation on galaxy luminosity and colour.
Absolute magnitudes are denoted $M_b$, $B$ and  $R$.
Distance dependent quantities are calculated using a Hubble parameter
$H=100\;Km\,s^{-1}\,Mpc^{-1}$.

\section{Analysis}

%===========================================================
\begin{table*}
\begin{center}

\label{tabla}
\begin{tabular*}{\textwidth}%
{|cc@{\extracolsep{\fill}}cc@{\extracolsep{\fill}}r|ccrrr|}

\hline
case& samples &restriction & $Log(\mathcal{M}_v/M_{\odot})$ & $r$ &
$N_1$ & $N_2$ &
$L_{KS}$  &  $P_{\beta}$  & $P_{\Delta K}$  \\
\hline \hline
\multicolumn{10}{|c|}{Spectral type segregation} \\ \hline

1 & $\eta<-1.4$ vs. $\eta>1.1$  & all luminosities  & all masses  & $<1.2$
& 8985 & 2561 & $>99.9\%$ & $<0.001$ & $<0.001$ \\

2 & $\eta<-1.4$ vs. $\eta>-1.4$ & all luminosities  & all masses  & $<1.2$
& 8985 & 6066 & $>99.9\%$ & $<0.001$ & $<0.001$ \\

3 & $\eta<-1.4$ vs. $\eta>1.1$ & all luminosities  & all masses  & $<0.5$
& 7234 & 1803 & $>99.9\%$ & $<0.001$ & $<0.001$ \\

4 & $\eta<-1.4$ vs. $\eta>-1.4$ & all luminosities  & all masses  & $<0.5$
& 7234 & 4250 & $>99.9\%$ & $<0.001$ & $<0.001$ \\

5 & $\eta<-1.4$ vs. $\eta>1.1$ & $M_b<-19$ & all masses  & $<0.5$
& 4086 & 603  & $>99.9\%$ & $0.001$ & $<0.001$ \\

6 & $\eta<-1.4$ vs. $\eta>1.1$ & $M_b>-19$ & all masses  & $<0.5$
& 3148 & 1200 & $>99.9\%$ & $0.001$ & $0.004$ \\

7 & $\eta<-1.4$ vs. $\eta>1.1$ & all luminosities  & $<14$  & $<0.5$
& 2430 & 916  & $99.9\%$ &$0.014$  & $0.001$ \\

8 & $\eta<-1.4$ vs. $\eta>1.1$ & all luminosities  & $>14$  & $<0.5$
& 4804 & 887  & $>99.9\%$ & $<0.001$ & $<0.001$ \\

\hline
\multicolumn{10}{|c|}{Luminosity segregation} \\ \hline

9 & $M_B<-19$ vs. $M_B>-19\;\;$ & all types  & all masses & $<0.5$
&5770 &5714 &$99.9\%$ &$0.001$ & $<0.001$ \\

10& $M_B<-19$ vs. $M_B>-19\;\;$ & all types  & all masses & $>0.5$
&1669 &1898 &$85.0\%$ &$0.104$ & $0.176$ \\

11& $M_B<-19$ vs. $M_B>-19\;\;$ & $\eta<-1.4$  &all masses & $<0.5$
&4086 &3148 &$99.0\%$ &$0.052$ & $0.007$  \\

12& $M_B<-19$ vs. $M_B>-19\;\;$ & $\eta>-1.4$  &all masses & $<0.5$
&1684 &2566 &$91.2\%$ &$0.018$ & $0.136$  \\

13& $M_B<-19$ vs. $M_B>-19\;\;$ & $\eta>1.1$  & all masses & $<0.5$
&603 &1200 &$54.0\%$ &$0.395$ & $0.280$ \\

14& $M_B<-19$ vs. $M_B>-19\;\;$ & all types  & $<14$  & $<0.5$
&1841 &2660 &$99.9\%$ &$<0.001$ & $<0.001$ \\

15& $M_B<-19$ vs. $M_B>-19\;\;$ & all types  & $>14$  & $<0.5$
&3929 &3054 &$94.1\%$ &$0.016$ & $0.019$  \\

\hline
\multicolumn{10}{|c|}{Colour index segregation} \\ \hline

16& $CI < 1 $ vs. $CI > 1 $ & all types \& luminosities &all masses & $<0.5$
&7291 &4180 &$>99.9\%$ &$<0.001$ & $<0.001$ \\

17& $CI < 1 $ vs. $CI > 1 $ & all types \& luminosities &all masses & $>0.5$
&1904 &1663 &$94.0\%$ &$0.276$ & $0.373$ \\

18& $CI < 1.2 $ vs. $CI > 1.2 $ & $\eta<-1.4$  &all masses& $<0.5$
&3583 &3651 & $90.5\%$ & $0.158$ & $0.239$\\

19& $CI < 0.7 $ vs. $CI > 0.7 $ & $\eta>1.1$   &all masses & $<0.5$
&805 &998 &$66.2\%$ &$0.369$ & $0.320$  \\

20& $CI < 1 $ vs. $CI > 1 $ & $\eta>1.1$   &all masses & $<0.5$
&136 &1666 &$45.3\%$ &$0.281$ & $0.197$  \\

21& $CI < 1 $ vs. $CI > 1 $ & $\eta<-1.4$   &all masses & $<0.5$
&6361 &866 &$99.8\%$ &$0.018$ & $<0.000$  \\

22& $CI < 1   $ vs. $CI > 1   $ & $M_b<-19$  &all masses & $<0.5$
&4121 &1639 & $>99.9\%$ & $0.002$ & $<0.001$\\

23& $CI < 1   $ vs. $CI > 1   $ & $M_b>-19$  &all masses & $<0.5$
&3170 &2541 & $>99.9\%$ &  $0.037$ & $<0.004$\\

24& $CI < 1 $ vs. $CI > 1 $ & all types \& luminosities & $<14$  & $<0.5$
&2341 &2154 &$>99.9\%$   &  $0.013$ & $<0.001$  \\

25& $CI < 1 $ vs. $CI > 1 $ & all types \& luminosities & $>14$  & $<0.5$
&4950 &2026 &$>99.9\%$ &$0.001$ & $<0.001$ \\

\hline
\end{tabular*}

\caption{Segregation effects in the 2dFGRS.  We list the results
for the three types of velocity segregations, analysed in this
work.  The different criteria used to divide the total sample are
shown.  Each case analysed has been labelled for its reference in
the text.  $N_1$ and $N_2$ are the number of galaxies in each
subsample; $P_{\beta}$ and $P_{\Delta K}$ are the probabilities
of obtaining a value from synthetic data, of $\beta$ or $\Delta
K$ respectively, greater than that obtained on the data.
$L_{KS}$ is the level of significance with wich the null
hypothesis that the two samples were taken from the same
distribution can be disproved, according to the
Kolmogorov--Smirnov test.} 

\end{center} 
\end{table*}
%===========================================================

\subsection{Ensemble group}

In order to make a suitable analysis of the data, we have
combined all the groups into a single system.  In this ensemble,
velocities $v$ and radius $r$ are expressed adimensionally.  The
line of sight velocities of each object $\Delta V$, relative to the
group average velocity, are scaled by the corresponding velocity
dispersion $\sigma$ of the host group. In a similar fashion,
galaxy projected distance to the group centre $R_p$ are scaled by
its virial radius value $R_v$.  This procedure allows for a
simple and improved statistical treatment of the data and,
assuming isotropy with respect to their centres, maintains
spatial and dynamical properties of the ensemble groups.
Similar procedures have been implemented e. g. by
\citet{adami,stein} and \citet{amelia}.

The virial radius is estimated from the projected distances of
members to the group centre \citep{manuelyz}. Uncertainties in these
value, as well as in the determination of the centre of the system
(derived by the unweighted mean of group member positions), can be
larger for groups with few galaxies. Accordingly we have restricted
all our analysis to groups with more than 10 members to reduce these
uncertainties. In section \ref{uncertain} we analyse the reliability
of our results related to this choice.

In order to explore the possible differences in the velocity
distribution of galaxies between groups and clusters, we have
divided the total sample of galaxies according to membership into
low ($<10^{14}\;M_{\odot}$) and high ($>10^{14}\;M_{\odot}$)
virial mass systems. In Fig. 1 we show the resulting velocity
distribution functions $f(v)$ where it can be appreciated that
both sets present remarkably similar distributions. This fact
allows us to perform the same treatment to all groups
irrespective of their mass.  We notice, however, that the shape
of this distribution is influenced by uncertainties in galaxy
redshift determinations in the 2dFGRS. The rms uncertainty is
approximately 85 km/s \citep{colles}. In section \ref{uncertain}
we explore the effect of these uncertainties in our analysis.

\subsection{Analysis of the ensemble group} \label{ensamble}

Our analysis is based on the comparison of normalized velocity
distributions of galaxies in different samples, selected by
spectral type ($\eta$), colour index (CI) or luminosity (quantified
by $M_b$).

In order to test the presence of a difference in the dynamical
behavior of two given samples of galaxies, we have adopted
different procedures to deal with the velocity distribution
functions $f(v)$.  One of these methods is the
Kolmogorov--Smirnov (KS) test \citep{press-numrec}.  In this
method, it is possible to disprove, to a given level of
significance, the null hypothesis that two distributions were
taken from the same population distribution function.

We have calculated the difference $\Delta K$ between the values
of the kurtosis of the two distribution functions. This parameter
provides a useful characterisation of the velocity distributions since
it is related to the relative fraction of galaxies with low velocity
($v<1$) in each subsample objectively.

We have also binned the velocities in $|v|$, using 5 bins. The
uncertainties in each bin have been determined using the bootstrap
resampling technique.  We defined a parameter $\beta$ as the
difference between the first bins of each distribution. The
uncertainty of $\beta$ is estimated by propagating individual errors
for each bin.

We have constructed 1000 new samples drawn from the original data
but reassigning spectral types, colour indexes or luminosities.
The distributions of these parameters mimics the ones of the
original observed sample. For each one of these random samples we
have determined $\beta$ and $\Delta K$. The resulting
distributions of these parameters provide an estimate of the
significance of the observed values in the real data, since these
distributions can be used to calculate the probabilities
$P_{\beta}$ and $P_{\Delta K}$ of obtaining, in the random
samples, values of $\beta$ and $\Delta K$ greater than the
observed ones.
The resulting probabilities $P_{\beta}$, $P_{\Delta K}$ and the
KS significance are shown in Table 1.

As previously mentioned, we have resampled sets of data using the
bootstrap technique in order to estimate the reliability of the
results. Explicitly, for each pair of samples we have calculated the
variance of $\beta$ and $\Delta K$ for the corresponding
bootstrapped samples. This variance provides a reliable estimate
of the uncertainties of the given parameter.

The large size of the data set has allowed us to explore the
results for different subsamples of the data corresponding to
different galaxy and group properties such as spectral types,
luminosities, colour indexes, group centric distance and parent
group virial mass. To achieve this goal we have applied the tests
described above to the different subsamples.

\section{results}

\begin{figure}
  \includegraphics[width=\columnwidth]{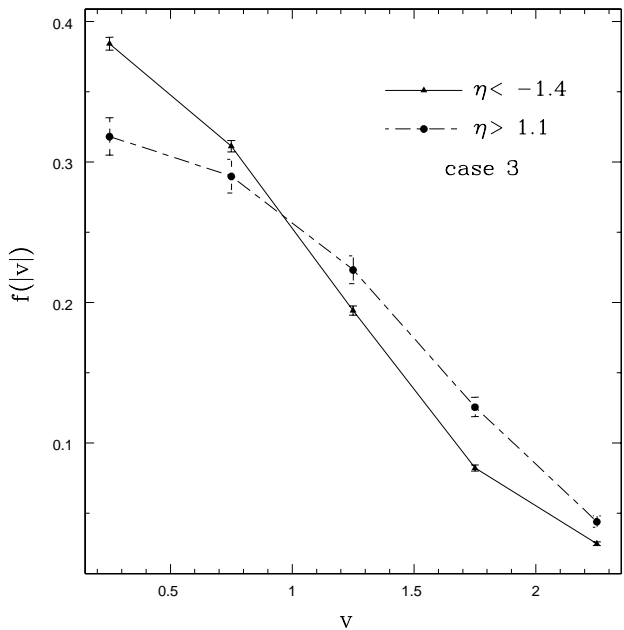}
  \label{STS01}
  \caption{Normalized binned velocity distributions for 
           early ($\eta<-1.4$) and late type ($\eta>1.1$) galaxies
           within $R_p=0.5\;R_v$.
	   Error bars have been calculated using the bootstrap
	   resampling method.  This plot corresponds to the
           case $3$ in Table 1, which presents the 
           strongest segregation.}
\end{figure}%
\begin{figure}
  \includegraphics[width=\columnwidth,trim=0 8.3cm 0 0,clip]%
                   {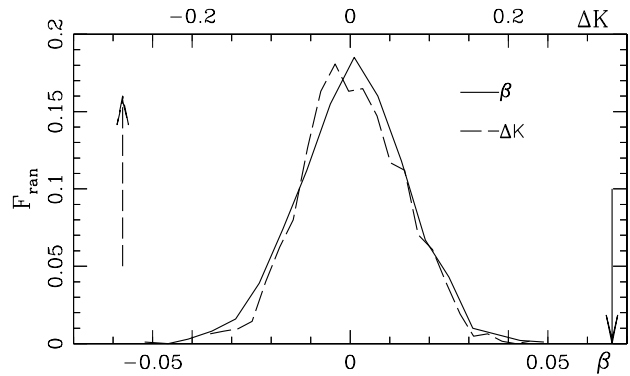}
  \label{errs}
  \caption{Distributions of $\beta$ and $\delta K$ parameters
           for 1000 random realizations for case 3
           (type I vs. type III-IV; in the inner region).
           Arrows indicate the observed values of the corresponding
           parameters.}
\end{figure}%

\subsection{Spectral type segregation} \label{STS}

\begin{figure}
  \includegraphics[width=\columnwidth]{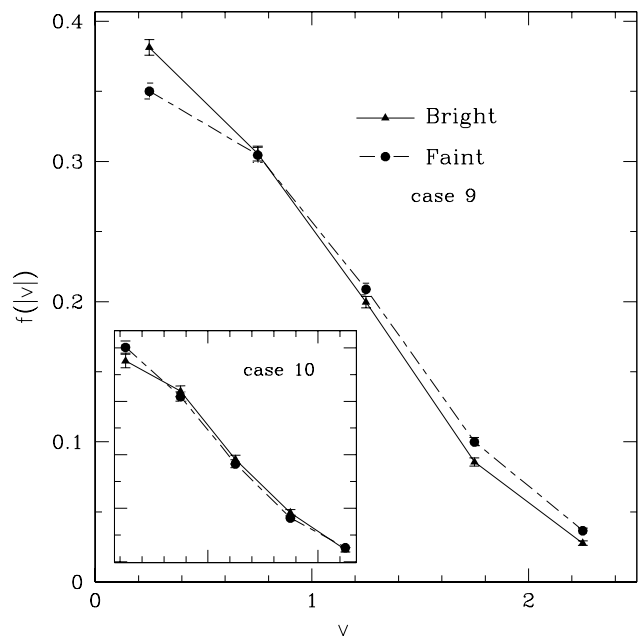}
  \label{LUM01}
  \caption{Normalized velocity distributions for bright 
	   (\mbox{$M_{b}<-19$}, solid line) and faint 
	   (\mbox{$M_{b}>-19$}, dashed line) galaxies within 
           \mbox{$r=0.5$} (case 9). The inner plot shows the results 
           obtained for a sample with $r>0.5$ (case 10)}
\end{figure}%

We have analysed the distribution functions of the relative velocities
for subsamples of different spectral type index, with no further
restriction in galaxy luminosity or colour index.
Figure 2 shows the velocity distributions $f(v)$ for
type I and types III--IV galaxies, where the
presence of a strong segregation can be clearly appreciated. The
relative line of sight velocities of type I galaxies are statistically
smaller than those of galaxies with a substantial star formation
activity.  To compute the statistical significance of the difference
between these two distributions, we have applied the three tests
described in the previous section. 
According to the KS test, we can disprove to a significance level
greater than 99.9\% that the two distributions are drawn
from the same parent distribution.
The distributions of the parameters $\beta$ and $\Delta K$
obtained for 1000 random realizations are shown in Fig. 3,
the arrows indicate the values obtained for the real data.

According to these results, velocity segregation by spectral type
is very significant statistically, corresponding to a fraction of
$69.5\%$ of early type galaxies and $60.7\%$ of late type galaxies 
within the group mean velocity dispersion ($v<1$) 
with respect to the total number of galaxies in each subsample.
These quantities may depend on the uncertainties on the
determination of velocities, however, as is stated in section
4.5, this fact does not affect the observed trends.

Table 1 summarizes the results for spectral type segregation
obtained by restricting the samples to a given galaxy luminosity
and group centric distance, as well as different parent group
virial mass.  If we restrict our analysis to the inner region of
the groups ($r<0.5$), the segregation intensity is stronger, but
it is similar for systems of different mass.  These facts will be
addressed in more detail in section \ref{globaldep}

\subsection{Luminosity segregation}

\begin{figure}
  \includegraphics[width=\columnwidth]{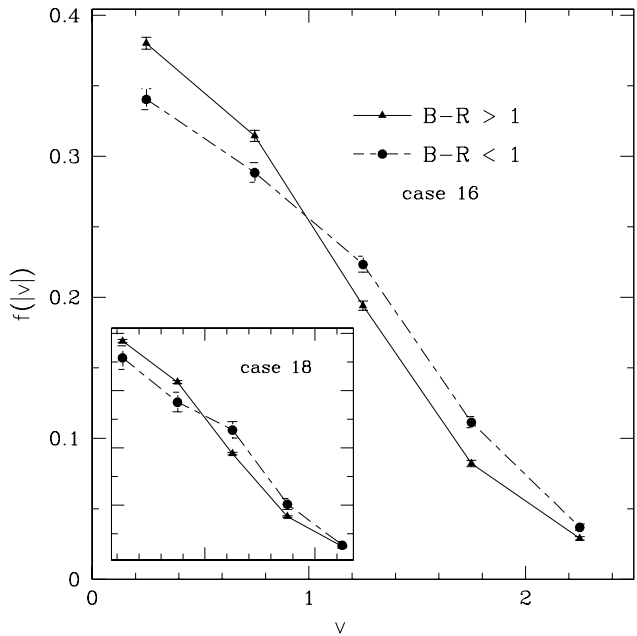}
  \label{CLI01}
  \caption{Normalized velocity
           distributions for red ($B\!-\!R\!>\!1$, solid line)
           and blue ($B\!-\!R\!<\!1$, dashed line) galaxies. The inner
	   plot shows the results obtained for a sample restricted to
           type I galaxies}
\end{figure}%

We have searched for a possible velocity segregation by
luminosity, by applying the same methods as in section \ref{STS}.
We find a significant difference of the normalized relative
velocity distributions between bright and faint galaxies, as can
be appreciated in Fig. 4.  The luminosity cut adopted to define
the two subsamples, $M_{b}=-19$, give a similar number of
galaxies in both of them.  We have also considered the central
($r < 0.5 $) region of the groups, where it can be seen that the
velocity segregation by luminosity is stronger (see Table 1).
This trend is similar to the observed behavior of the segregation
by spectral type, which is stronger in denser environments (see
section \ref{STS}) and will be discussed in more detail in
section \ref{globaldep}.

Given the significant velocity segregation by spectral types, we
have searched for luminosity segregation in subsamples restricted
to type I and types III-IV galaxies. We find that the luminosity
segregation signal is of less significance in these cases,
indicating that luminosity is not a primary parameter in defining
the dynamics of galaxies in groups.  This suggests that early
spectral type galaxies, on average more luminous than late types,
could provide the observed dependence on luminosity.

\subsection{Colour segregation}

Using the same procedure we have also explored the relation
between galaxy dynamics and colour index. We have considered the
$B\!-\!R$ colour index provided in the 2dFGRS final data release
for the galaxies in the group sample. Given the narrow range of
redshifts in the group sample
($0.02\,\lesssim\,z\,\lesssim\,0.20$), a unique threshold is
suitable to define a sample of red galaxies.

The normalized velocity distributions for red ($B\!-\!R\!>\!1$) and
blue ($B\!-\!R\!<\!1$) galaxies are shown in Fig. 5, where a
significant velocity segregation of velocity according to galaxy
colour index can be clearly appreciated. Given the correlation between
$B\!-\!R$ colour index and spectral index $\eta$, we have restricted
our analysis to galaxies with low present star formation (type I), and
strongly star forming galaxies (types III-IV). The results are shown
in Table 1, and for type I objects in the small box of Fig. 5.  As it
can be appreciated, velocity segregation has a lower level of
significance in the last case.

\subsection{Dependence of segregation on global properties}
\label{globaldep}

\begin{figure}
  \includegraphics[width=\columnwidth]{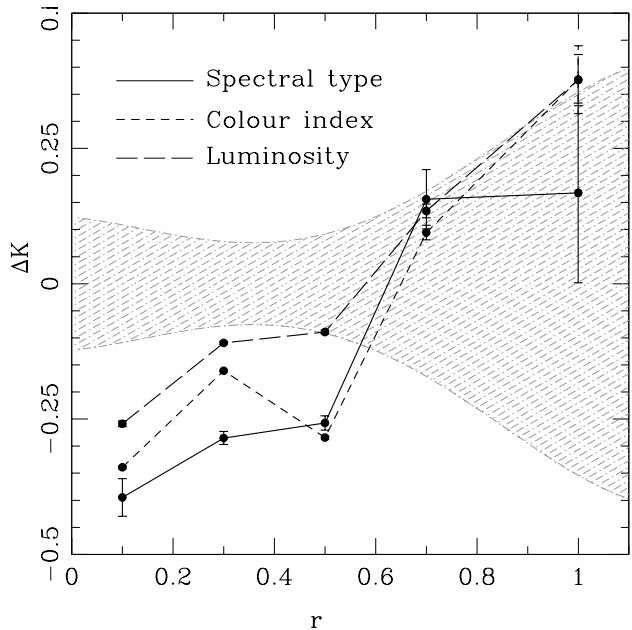}
  \label{STS.rad}
  \caption{Dependence of the kurtosis differences $\Delta K$
  	   on $R_p/R_{vir}$. The dashed region displays the rms of the
           distribution of $\Delta K$ obtained for the random samples
           described in section \ref{ensamble}. 
	   Error bands are calculated using bootstrap resampling technique.}
\end{figure}%

\begin{figure}
  \includegraphics[width=\columnwidth]{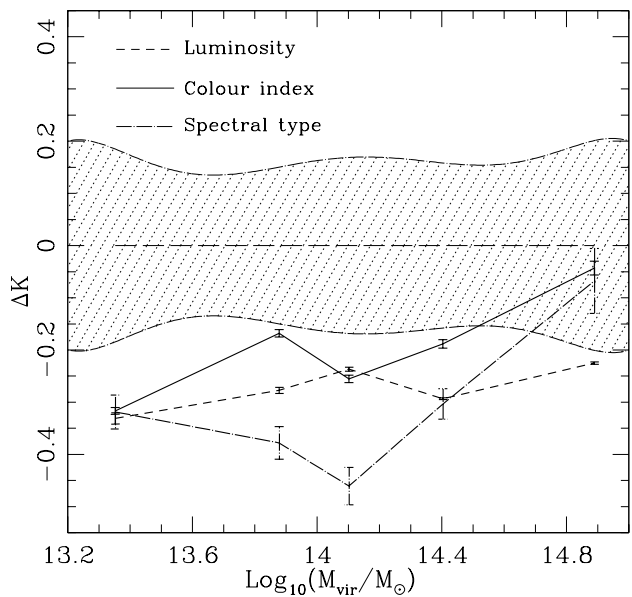}
  \label{STS+CI.mass}
  \caption{The same as Fig. 6 but for the dependence of $\Delta K$
  	   on the virial mass of the parent group.  The dashed
           region displays the rms of the distribution of 
           $\Delta K$ obtained for the random samples.
	   Error bands are calculated using bootstrap resampling technique.}
\end{figure}%

In the previous section we have analysed the different dynamics
of galaxies according to their intrinsic properties, namely
spectral type, luminosity and colour index. Our results suggest
that the segregation effects are stronger in denser environments
although there is no indication of a strong dependence on the
parent group mass. In this section we explore in more detail the
dependence of our results on galaxy--group centric distance and
parent group virial mass.

In Fig. 6 we show the dependence of the kurtosis differences
$\Delta K$ on $r$. The dashed region displays the rms of the
distribution of $\Delta K$ obtained for the random samples
described in section \ref{ensamble}. It can be clearly
appreciated that velocity segregation in the central regions ($r<0.5$) is
particularly important for spectral type and colour index but
smoothly decreases at larger group centric distances.  In a similar
fashion, we have analysed the dependence of velocity segregation
on parent group virial mass.  Our results for galaxies with
$r<0.5$ are shown in Fig. 7.  It can be appreciated that there is
no strong dependence of the segregation effects on the mass of
the parent group.  
This is an important fact suggesting that the
mechanisms that generate the observed difference in the
dynamics, according to the galaxy star formation activity, 
are efficient on a wide range of masses.

\subsection{Analysis of uncertainties}
\label{uncertain}

The uncertainty in redshift determinations in the 2dFGRS amounts
to a rms of $85Km\;s^{-1}$ \citep{colles}. This is a quite large
figure, so its effects on our statistical analysis deserves
particular attention.  To account for this uncertainty we have
convolved each line of sight velocity measurement with a Gaussian
with dispersion $\epsilon=85Km\;s^{-1}$. The resulting smoothed
histograms are suitable to compute the parameters characterizing
the differences of the distribution of velocities of two given
samples of galaxies taking into account line of sight velocity
errors. As an example of this analysis, in Fig. 8 we show the
results for case 3, where it can be appreciated that the 
binned and the smoothed distributions
show the same behavior.  Moreover
we have computed the parameters $\beta$ and $\Delta K$ for the
cases shown in Table 1, finding similar results which show the
stability of our analysis against redshift measurements errors.

As a test of the stability of our results against the number of
group members, we have also analysed a sub-sample of groups
restricted to have at least 20 members.  We obtain similar and
even more prominent segregation effects for the same galaxy
properties analysed previously.  Also, in order to test the
effects of possible erroneous determination of the centre of the
groups in our results, we have repeated the analysis for
re--centered groups. These new centres where calculated using
only galaxies within 1.2 times the virial radius and 2.5 times
the velocity dispersion, which would provide a better estimate of
group centres for elongated or clumpy systems. Again here the
results are similar and show that our conclusions are not
strongly dependent on the group centre definition.

\section{Discussion}

\begin{figure}
  \includegraphics[width=\columnwidth]{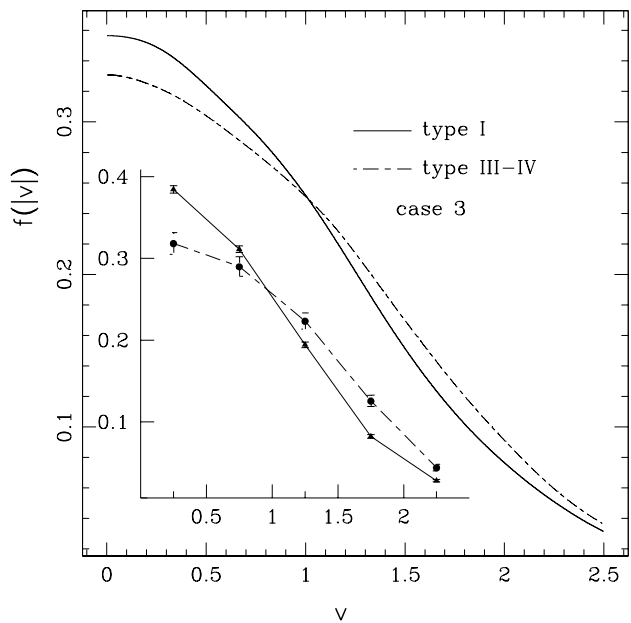}
  \label{smooth}
  \caption{Normalized velocity distributions convolved with a Gaussian
           of width $85\;Km\,s^{-1}$.  Inside box shows the binned case
           for comparison.}
\end{figure}%

We find a statistically significant difference of the
distributions of group centric line of sight velocities,
normalized to the group mean velocity dispersion, for samples of
galaxies selected by spectral type, luminosity or colour index.
Given the large data set analysed, we have been able to
investigate the dependence of this velocity segregation on group
properties and galaxy--group centric distance.  Spectral type I
objects, corresponding to passively star forming galaxies, show a
statistically narrower velocity distribution than that of
galaxies with a substantial star formation activity (types
III-IV). Similarly, samples of galaxies with greater colour index
($B\!-\!R\!>\!1$) have a larger fraction of small velocities
($v<1$) compared to galaxies with $B\!-\!R\!<\!1$.  These two
trends show a strong correlation between galaxy dynamics in
groups and star formation, reflected both by spectral type
and by colour index.

The velocity distribution of luminous galaxies (typically
brighter than $M_b=-19$) also show a larger fraction of small
velocities, although we notice that once the galaxies are
restricted to a given spectral type, there is a less significant
segregation. Thus, luminosity is not likely to be a primary
parameter determining galaxy dynamics in groups. Our results
suggest that the observed luminosity segregation might be related
to the fact that the slowest objects, of early spectral type, are
on average more luminous than star forming galaxies.

There are several mechanisms that may produce dynamical
segregations of galaxies in groups and clusters. 
Ram pressure can effectively remove the existing gas in the
galaxies and transform star forming into passively star forming
objects.  This mechanism affects most strongly those galaxies
with large velocities with respect to the intra--cluster medium,
and then it should produce a dynamical segregation with opposite
trends to the observed one.  Moreover, since our analysis
concerns groups and small clusters of galaxies, ram pressure is
not expected to be significant.  Our results indicates that the
velocity segregation effects are nearly independent of group
virial mass.  This fact also suggests that ram pressure is not
important since its effects are stronger in more massive systems,
and with higher velocity dispersion. Thus, it is
unlikely that ram pressure may explain the observed correlations. 

Mergers on the other hand, are effective to generate spheroidal
objects with a low star formation rate.  Galaxy encounters are
expected to lower the original cluster-centric relative
velocities of each galaxy, with respect to the velocities of the
galaxies prior to the merger event, so that they can act
effectively in generating the observed trends.

In a similar fashion, tidal interactions may effectively remove a
substantial amount of gas from disks of galaxies, and then, are
also effective in truncating star formation.  Early type objects
generated through this mechanism, would be biased to smaller
velocities since interactions are expected to be more effective
in slow encounters.  Furthermore, these are generally brighter
and redder objects.

It has been suggested that morphological transformation of galaxies
takes place in systems which have a density threshold larger than the
density of groups and poor clusters of galaxies \citep{moore, gray}.
Our results indicate that dynamical segregation of passively star
forming galaxies is a generic feature of systems of galaxies,
irrespective of global properties. However, the fact that segregation
effectively occurs in the inner regions of groups indicates that
density might be an important parameter in determining the observed
effects.  

\section*{Acknowledgments}

This work was partially supported by the Concejo Nacional de 
Investigaciones Cient\'{\i}ficas y Tecnol\'ogicas (CONICET), 
the Secretar\'{\i}a de Ciencia y T\'ecnica (UNC) and 
the Agencia C\'ordoba Ciencia.

\end{document}